\documentclass[12pt,amsbsy]{article}
\usepackage{amsmath}
\usepackage{amssymb}
\usepackage{graphicx}
\usepackage{psfig}

\newcommand{\beq}{\begin{equation}}
\newcommand{\eeq}{\end{equation}}
\newcommand{\beqn}{\begin{eqnarray}}
\newcommand{\eeqn}{\end{eqnarray}}

\newcommand{\lsim}{\mbox{$<$\hspace{-0.8em}\raisebox{-0.4em}{$\sim$}}}

\newcommand{\al}{\mbox{${\alpha}$}}
\newcommand{\be}{\mbox{${\beta}$}}
\newcommand{\ga}{\mbox{${\gamma}$}}
\newcommand{\Ga}{\mbox{${\Gamma}$}}

\newcommand{\De}{\mbox{${\Delta}$}}

\newcommand{\om}{\mbox{${\omega}$}}

\newcommand{\pa}{\mbox{${\partial}$}}

\begin{document}

\begin{titlepage}

\vspace{1cm}

\begin{center}
{\large Radiation of Quantized Black Hole}
\end{center}

\begin{center}
I.B. Khriplovich\footnote{khriplovich@inp.nsk.su}
\end{center}
\begin{center}
Budker Institute of Nuclear Physics\\
630090 Novosibirsk, Russia,\\
and Novosibirsk University
\end{center}

\bigskip

\begin{abstract}
The maximum entropy of a quantized surface is demonstrated to be
proportional to the surface area in the classical limit. The
general structure of the horizon spectrum and the value of the
Barbero-Immirzi parameter are found. The discrete spectrum of
thermal radiation of a black hole fits naturally the Wien profile.
The natural widths of the lines are very small as compared to the
distances between them. The total intensity of the thermal
radiation is calculated.
\end{abstract}

\vspace{9cm}

\end{titlepage}

\section{Introduction}

The idea of quantizing the horizon area of black holes was put
forward many years ago by Bekenstein in the pioneering
article~\cite{bek}. He pointed out that reversible transformations
of the horizon area of a nonextremal black hole found by
Christodoulou and Ruffini~\cite{ch,chr} have an adiabatic nature.
Of course, the quantization of an adiabatic invariant is perfectly
natural, in accordance with the correspondence principle.

Once this hypothesis is accepted, the general structure of the
quantization condition for large quantum numbers gets obvious, up to
an overall numerical constant $\be$. The quantization condition for
the horizon area $A$ should be
\begin{equation}\label{qu}
A=\be \, l^2_p \, N,
\end{equation}
where $N$ is some large quantum number~\cite{kh}. Indeed, the
presence of the Planck length squared $l^2_p = k \hbar /c^3$ is
only natural in this quantization rule. Then, for the horizon area
$A$ to be finite in the classical limit, the power of $N$ should
be the same as that of $\hbar$ in $l^2_p$. This argument can be
checked by considering any expectation value in quantum mechanics,
nonvanishing in the classical limit. It is worth mentioning that
there are no compelling reasons to believe that $N$ is an integer.
Neither there are compelling reasons to believe that the spectrum
(\ref{qu}) is equidistant~\cite{kh1,kh2}.

On the other hand, formula (\ref{qu}) can be interpreted as
follows. The whole horizon area $A$ is split into elements of
typical size $\sim l_p^2\,$, each of them giving a contribution to
the large quantum number $N$. This scheme arises in particular in
the framework of loop quantum gravity (LQG)~[7--11].

A quantized surface in LQG looks as follows. One ascribes to it a
set of edges. Each edge is supplied with an integer or
half-integer ``angular momentum'' $j$:
\beq\label{j}
j= 1/2, 1, 3/2, ...\; .
\eeq
The projections $m$ of these ``angular momenta'' run as usual from
$-j$ to $j$. The area of a surface is
\beq\label{Aj}
A =8\pi\ga\, l_p^2 \sum_i \sqrt{j_i(j_i+1)}\,.
\eeq
The numerical factor $\ga$ in (\ref{Aj}) cannot be determined
without an additional physical input. This free (so-called
Barbero-Immirzi) parameter \cite{imm,rot} corresponds to a family
of inequivalent quantum theories, all of them being viable without
such an input.

We mention that though the spectrum (\ref{Aj}) is not equidistant,
it is not far away from it. Indeed, even for the smallest quantum
number $j=1/2$, $\sqrt{j(j+1)}$ can be approximated by $j+1/2$
with an accuracy 13\%. And the approximation $\sqrt{j(j+1)}
\approx j+1/2$ gets better and better with growing $j$, i.e. the
spectrum (\ref{Aj}) approaches an equidistant one more and more.
This feature of the spectrum (\ref{Aj}) is of interest in
connection with the observation due to Bekenstein: quantum effects
result in the following lower bound on the change of the horizon
area $\De A$ under an adiabatic process:
\beq\label{da}
(\De A)_{min}=\xi l_p^2\,;
\eeq
here $\xi$ is a numerical factor reflecting ``the inherent
fuzziness of the uncertainty relation''~\cite{bek1}. Of course,
right-hand-side of formula (\ref{da}) is proportional to $\hbar$,
together with the Planck length squared $l_p^2$ .

Due to the uncertainty of the numerical factor $\xi$ itself, one
cannot see any reason why $\xi$ should not slightly change from
one act of capture to another. So, the discussed quasiequidistant
spectrum (\ref{Aj}) agrees with the bound (\ref{da}), practically
as well as the equidistant one. Below we will come back to
relation (\ref{da}).

As to the unknown parameter $\ga$ in (\ref{Aj}), the first
attempts to fix its value, based on the analysis of the black hole
entropy, were made in~\cite{rov,kra}. However, these attempts did
not lead to concrete quantitative results.

Then it was argued in~\cite{asht} that for the black hole horizon
all quantum numbers $j$ are equal to $1/2$ (as it is the case in
the so-called ``it from bit'' model formulated earlier by
Wheeler~\cite{whe}). With these quantum numbers, one arrives
easily at the equidistant area spectrum and at the value $\ga =
\ln 2/(\pi \sqrt{3})$ for the Barbero-Immirzi parameter. However,
the result of~\cite{asht} was demonstrated in ~\cite{kh1} to be
certainly incorrect\footnote{Later, the result of \cite{asht} was
criticized as well in~\cite{lew,mei}. Then an error made
in~\cite{asht} was acknowledged~\cite{asht1}. We will demonstrate
below that the result of~\cite{lew,mei} is also incorrect.} since
it violates the so-called holographic bound formulated
in~[22--24]. According to this bound, among the spherical surfaces
of a given area, it is the surface of a black hole horizon that
has the largest entropy.

\section{Microcanonical Entropy of Black Hole}

On the other hand, this requirement of maximum entropy allows one
to find the correct structure of the horizon area~\cite{kk}, which
in particular is of crucial importance for the problem of
radiation of a quantized black hole.

We consider in fact the ``microcanonical'' entropy $S$ of a
quantized surface defined as the logarithm of the number of states
of this surface for a fixed area $A$ (instead of fixed energy in
common problems). Obviously, this number of states $K$ depends on
the assumptions concerning the distinguishability of the edges.

To analyze the problem, it is convenient to rewrite formula
(\ref{Aj}) as follows:
\beq\label{Anu}
A = 8\pi\ga\, l_p^2 \sum_{jm}\sqrt{j(j+1)}\;\nu_{jm}\,.
\eeq
Here $\nu_{jm}$ is the number of edges with given $j$ and $m$. It
can be demonstrated~\cite{kh1,kh2} that the only reasonable
assumption on the distinguishability of edges that may result in
acceptable physical predictions (i.e. may comply both with the
Bekenstein-Hawking relation and with the holographic bound) is as
follows:\\

\begin{tabular}[h]{cccc}
 \vspace{3mm}
 nonequal $j$, & any $m$ & $\longrightarrow$ &
distinguishable;\\
\vspace{3mm}
equal $j$, & nonequal $m$ &
$\longrightarrow$ & distinguishable;\\
 \vspace{3mm}
equal $j$, & equal $m$ & $\longrightarrow$ & indistinguishable.\\
\end{tabular}\\

Under this assumption, the number of states of the horizon surface
for a given number $\nu_{jm}$ of edges with momenta $j$ and their
projections $j_z=m$, is obviously
\beq\label{mk}
K = \nu\,!\, \prod_{jm}\,\frac{1}{\nu_{jm}\,!}\;, \quad {\rm
where} \quad \nu =\sum_j \nu_j\,, \quad \nu_j = \sum_m \nu_{jm}\,,
\eeq
and the corresponding entropy equals
\beq\label{ms}
S=\ln K = \ln(\nu\,!)\,- \sum_{jm}\,\ln(\nu_{jm}\,!)\,.
\eeq
The structures of the last expression and of formula (\ref{Anu}) are
so different that in a general case the entropy certainly cannot be
proportional to the area. However, this is the case for the maximum
entropy in the classical limit.

In this limit, with all effective ``occupation numbers'' large,
$\nu_{jm} \gg 1$, we use the Stirling approximation so that the
entropy is
\beq\label{en2}
S= \nu \ln \nu -\sum_{jm} \nu_{jm} \ln \nu_{jm}\,.
\eeq
We calculate its maximum for a fixed area $A$, i.e. for a fixed sum
\beq\label{N}
N\,=  \sum_{jm}^\infty \sqrt{j(j+1)}\,\nu_{jm}={\rm const} \,.
\eeq

The problem reduces to the solution of the system of equations
\beq\label{sys}
\ln \nu  - \ln \nu_{jm} = \mu \sqrt{j(j+1)}\,,
\eeq
where $\mu$ is the Lagrange multiplier for the constraining relation
(\ref{N}). These equations can be rewritten as
\beq\label{nu1}
\nu_{jm}=\nu e^{- \mu \sqrt{j(j+1)}},
\eeq
or
\beq\label{nu2}
\nu_j = (2j+1)e^{- \mu \sqrt{j(j+1)}}\nu.
\eeq
Now we sum expressions (\ref{nu2}) over $j$, and with $\sum_j \nu_j
= \nu$ arrive at the equation for $\mu$:
\beq\label{equ}
\sum_{j=1/2}^{\infty} (2j+1)\, e^{- \mu \sqrt{j(j+1)}} = 1.
\eeq
Its solution is
\beq\label{mu}
\mu = 1.722.
\eeq

Strictly speaking, the summation in formula (\ref{mu}) extends not
to infinity, but to some $j_{max}$. Its value follows from the
obvious condition: none of $\nu_{jm}$ should be less than unity.
Then, for $\nu \gg 1$ equation (\ref{nu1}) gives
\beq\label{jm}
j_{max} = \,\frac{\ln \nu}{\mu}\,.
\eeq
It is well-known that the Stirling approximation for $n!$ has
reasonably good numerical accuracy  even for $n = 1$. Due to it,
formula (\ref{jm}) for $j_{max}$ is not just an estimate, but has
reasonably good numerical accuracy. The relative magnitude of the
corresponding correction to (\ref{mu}) can be easily estimated as
$\sim \ln \nu/\nu$.

Let us come back now to equation (\ref{sys}). When multiplying it
by $\nu_{jm}$ and summing over $jm$, we arrive with the constraint
(\ref{N}) at the following result for the maximum entropy for a
given value of $N$:
\beq\label{enf0}
S_{\rm max}= 1.722\,N\,,
\eeq
so that with the Bekenstein-Hawking relation and formula (\ref{Anu})
we find the value of the Barbero-Immirzi parameter:
\beq\label{bip}
\ga = 0.274.
\eeq

Quite recently this calculation with the same result, though with
somewhat different motivation, was reproduced in~\cite{gm}.

It should be emphasized that the above calculation is not special
for LQG only, but applies (with obvious modifications) to a more
general class of approaches to the quantization of surfaces. The
following assumption is really necessary here: the surface should
consist of sites of different sorts, so that there are $\nu_i$
sites of each sort $i$, with a generalized effective quantum
number $r_i$ (here $\sqrt{j(j+1)}$), and a statistical weight
$g_i$ (here $2j+1$). Then in the classical limit, with given
functions $r_i$ and $g_i$ the maximum entropy of a surface can be
found, at least numerically, and it is certainly proportional to
the area of the surface.

As to previous attempts to calculate $\ga$, one should indicate an
apparent error in state counting made in~\cite{lew,mei}. It can be
easily checked that the transition from formula (25) to formulae
(29), (36) of~\cite{lew} performed therein and then employed
in~\cite{mei}, is certainly valid under the assumption that for
each quantum number $j$ only two maximum projections $\pm j$ are
allowed. But therefore it cannot hold for the correct number
$2j+1$ of the projections. No wonder that the equation for the BI
parameter in~\cite{mei} looks as
\beq\label{2}
2\sum_{j=1/2}^{\infty}\, e^{- \mu \sqrt{j(j+1)}} = 1\,,
\eeq
instead of ours (\ref{equ}) (see also the discussion of (\ref{2})
in~\cite{gm}).

The conclusion is obvious. Any restriction on the number of
admissible states for the horizon, as compared to a generic
quantized surface, be it the restriction to
\[
j=1/2\,, \quad m=\pm 1/2\,,
\]
made in~\cite{asht}, or the restriction to
\[
{\rm any} j\,, \quad m=\pm j\,,
\]
made in~\cite{lew,mei}, results in a conflict with the holographic
bound.

\section{Quantization of Rotating Black Hole}

When discussing the radiation spectrum of quantized black holes,
one should take into account the selection rules for angular
momentum. Obviously, radiation of any particle with nonvanishing
spin is impossible if both initial and final states of a black
hole are spherically symmetric. Therefore, to find the radiation
spectrum, the quantization rule for the mass of a Schwarzschild
black hole should be generalized to that of a rotating Kerr black
hole.

To derive the quantization rule for Kerr black hole, we come back
to the thought experiment analyzed in~\cite{ch,chr}. Therein,
under adiabatic capture of a particle with an angular momentum
$j$, the angular momentum $J$ of a rotating black hole changes by
a finite amount $j$, but the horizon area $A$ does not change. Of
course, under some other variation of parameters it is the angular
momentum $J$ that remains constant. In other words, we have here
two independent adiabatic invariants, $A$ and $J$, for a Kerr
black hole with a mass $M$.

Such a situation is quite common in ordinary mechanics. For
instance, the energy of a particle with mass $m$ bound in the
Coulomb field $U(r) = - \al/r$ is
\beq\label{Ec}
E=-\,\frac{m \al^2}{2\,(I_r + I_{\phi})^2}\,,
\eeq
where $I_r$ and $I_{\phi}$ are adiabatic invariants for the radial
and angular degree of freedom, respectively. Of course, the energy
$E$ is in a sense an adiabatic invariant also, but it is invariant
only with respect to those variations of parameters under which
both $I_r$ and $I_{\phi}$ remain constant. As to quantum
mechanics, in it formula (\ref{Ec}) goes into
\beq\label{Eq}
E=-\,\frac{m \al^2}{2\,\hbar^2\, (n_r + 1 + l)^2}\,,
\eeq
where $n_r$ and $l$ are the radial and orbital quantum numbers,
respectively.

This example prompts the solution of the quantization problem for
a Kerr black hole. It is conveniently formulated in terms of the
so-called irreducible mass $M_{ir}$ of a black hole, related by
definition to its horizon radius $r_h$ and area $A$ as follows:
\beq\label{rel}
r_h= 2 k M_{ir}\,, \quad A = 16\pi k^2 M_{ir}^2\,.
\eeq
Together with the horizon area $A$, the irreducible mass is an
adiabatic invariant. In accordance with (\ref{Aj}) and (\ref{N}),
it is quantized as follows:
\beq\label{mirq}
M_{ir}^2 =\,\frac{1}{2}\,m^2_p \,N\,,
\eeq
where $m^2_p=\hbar c/k$ is the Planck mass squared.

Of course, for a Schwarzschild black hole $M_{ir}$ coincides with
its ordinary mass $M$. However, for a Kerr black hole the
situation is more interesting. Here
\beq\label{mik}
M^2=M_{ir}^2 + \,\frac{J^2}{r_h^2}\,=M_{ir}^2 + \,\frac{J^2}{4k^2
M_{ir}^2}\,,
\eeq
where $J$ is the internal angular momentum of a rotating black
hole.

Now, with the account for equation (\ref{mirq}), we arrive at the
following quantization rule for the mass squared $M^2$ of a
rotating black hole:
\beq\label{mqj}
M^2 =\,\frac{1}{2}\,m^2_p \left[\ga N +\,\frac{J(J+1)}{\ga
N}\right].
\eeq
Obviously, as long as a black hole is far away from an extremal
one, i.e. while $\ga N \gg J$, one can neglect the dependence of
$M^2$ on $J$, and the angular momentum selection rules have
practically no influence on the radiation spectrum of a black
hole.

As to the mass and irreducible mass of a charged black hole, they
are related as follows:
\beq\label{mir}
M=M_{ir} + \,\frac{q^2}{2r_h}\,;
\eeq
here $q$ is the charge of the black hole. This formula has a
simple physical interpretation: the total mass (or total energy)
$M$ of a charged black hole consists of its irreducible mass
$M_{ir}$ and of the energy $q^2/2r_h$ of its electric field in the
outer space $r>r_h$.

With $r_h= 2 k M_{ir}$, relation (\ref{mir}) can be rewritten as
\beq\label{mic}
M^2=M_{ir}^2 + \,\frac{q^4}{16 k^2 M_{ir}^2} + \frac{q^2}{2k}\,.
\eeq
Thus, for a charged black hole $M^2$ is quantized as follows:
\beq\label{mqq}
M^2=\,\frac{1}{2}\,m^2_p \left[\ga N +\,\frac{q^4}{4\ga N}+ q^2
\right].
\eeq

In fact, relations of this type (even in a more general form, for
Kerr-Newman black holes, both charged and rotating) were presented
already in the pioneering article~\cite{bek}, though with the
equidistant quantization rule for $M_{ir}^2$, i.e. for the horizon
area (see also~\cite{bek1}). More recently, the conclusion that
the mass of a quantized black hole should be expressed via its
quantized area and angular momentum, was made in the approach
based on the notion of so-called isolated horizons~\cite{af,abl}.

I do not mention here those attempts to quantize rotating and
charged black holes which resulted in weird quantization rules for
$\hat{J}^2$ and $e^2/\hbar c$.

\section{Radiation Spectrum of Quantized Black Hole}

It follows from expression (\ref{mqj}) that for a rotating black
hole the radiation frequency $\om$, which coincides with the loss
$\De M$ of the black hole mass, is
\beq\label{om}
\om = \De M = T \mu \,\De N +\,\frac{1}{4kM}\,\frac{2J+1}{\ga N}\,
\De J\,,
\eeq
where $\De N$ and $\De J$ are the losses of the area quantum
number $N$ and of the angular momentum $J$, respectively. We have
used here, in line with (\ref{mqj}), the following identity for
the Hawking temperature $T$:
\beq
T=\,\frac{\pa M}{\pa S}\,=\,\frac{1}{8\pi k M}\,\frac{\pa M^2}{\pa
M_{ir}^2}\,,
\eeq
as well as formula (\ref{mik}).

In the same way, for a charged black hole one obtains with formula
(\ref{mqq}) the radiation frequency
\beq\label{omc}
\om = \De M = T \mu \,\De N +\,\frac{1}{4 k M}\,\left(2 +
\frac{q^2}{\ga N}\right)q\,\De q\,,
\eeq
where $\De q$ is the loss of the charge.

We will be interested mainly in the first, temperature terms in
(\ref{om}) and (\ref{omc}), dominating everywhere but the vicinity
of the extremal regime, where $J \to \ga N$, or $q^2 \to 2 \ga N$,
and $T \to 0$. The natural assumption is that the temperature
radiation occurs when an edge with a given value of $j$
disappears, which means that
\beq\label{dom}
\De N_j = r_j\,, \quad \om_j = T \mu\, r_j\,.
\eeq
Thus we arrive at the discrete spectrum with a finite number of
lines. Their frequencies start at $\om_{min}=T\mu\sqrt{3}/2$ and
terminate at $\om_{max}=T\ln\nu$. We recall here that $j \leq
j_{max} =\ln \nu/\mu$, so that the number of lines is not so
large, $\sim 10^2$, if the mass of black hole is comparable to
that of the Sun. However, due to the exponential decrease of the
radiation intensity with $\om$ or $j$ (see below), the existence
of $\om_{max}$ and finite number of lines are not of much
importance.

To substantiate the made assumption, we come back to the lower
bound (\ref{da}) on the change of the horizon area under an
adiabatic capture of a particle. The presence of the gap
(\ref{da}) in this process means that this threshold capture
effectively consists in the increase by unity of the occupation
number $\nu_{jm}$ with the smallest $j$, equal to $1/2$. If the
capture were accompanied by a reshuffle of few occupation numbers,
the change of the area could be easily made as small as one
wishes. For instance, one could delete two edges with quantum
numbers $j_1$ and $j_2$, and add an edge with a quantum number
$j_1+j_2$. Obviously, with $j_{1,2} \gg 1$ the area increase could
be made arbitrarily small.

It is only natural to assume that in the radiation process as
well, changing few occupation numbers instead of one is at least
strongly suppressed. In this way we arrive at
equations~(\ref{dom}).

Our next assumption, at least as natural as this one, is that the
probability of radiation of a quantum with frequency $\om_j$ is
proportional to the occupation number $\nu_j$. Correspondingly,
the radiation intensity $I_j$ at this frequency $\om_j$ is
proportional to $\nu_j\, \om_j$:
\beq\label{ij}
I_j \sim \nu_j\, \om_j \sim \nu\, (2j+1)\, \om_j \,
\exp(-\om_j/T)\,.
\eeq

Let us compare this expression with the intensity of the
black-body radiation in the Wien limit $\om/T \gg 1$,
\beq\label{w}
I(\om) = A\,\frac{\om^3}{4 \pi^2}\,\exp(-\om/T)d\om\,,
\eeq
where $A$ is the area of a spherical black body. First of all, our
relation (\ref{ij}) for $I_j$ reproduces directly the exponential
factor of the Wien spectrum. Then, $d\om$ in (\ref{w}) goes over
into $1/2\,\mu T$ since the limit $\om/T \gg 1$ corresponds in our
problem to $\sqrt{j(j+1)} \gg 1$, i.e. to $\sqrt{j(j+1)} \simeq
j+1/2$, and the minimum increment of $j$ is 1/2. Now, to reproduce
the Wien profile, we supplement relation (\ref{ij}) with the
following factors: some ``oscillator strength'' proportional to
$\om_j$, obvious powers of $\mu T$, the Newton constant $k$
(necessary to transform $\nu$ into $A$), and obvious numerical
ones. Thus we arrive at the final formula for the discrete
radiation spectrum of a black hole:
\beq\label{ija}
I_j =
A\,T^4\,\frac{\mu^4}{8\pi^2}\,j(j+1/2)(j+1)\exp\left(-\mu\sqrt{j(j+1)}\right)\,.
\eeq

Of course, since the Wien spectrum (\ref{w}) corresponds to $j\gg
1$, one cannot guarantee the exact structure of $j$-dependence in
formula (\ref{ija}), especially in the preexponential factor. For
instance, it would be perhaps as legitimate to write therein
$j^{3/2}(j+1)^{3/2}$ instead of $j(j+1/2)(j+1)$. However, this
ambiguity is not as essential, at least numerically.

We note that since the black hole temperature $T$ is less than the
minimum allowed frequency $\om_{min}$, this spectrum has no
Rayleigh-Jeans region at all.

Now, the emission probability for a quantum of frequency $\om_j =
T \mu\, r_j$, i.e. the width of the corresponding line, is
\beq\label{lw}
\Ga_j = \,\frac{I_j}{\om_j}\,=
A\,T^3\,\frac{\mu^3}{8\pi^2}\,(j+1/2)\sqrt{j(j+1)}\exp\left(-\mu\sqrt{j(j+1)}\right)\,.
\eeq
The ratio of this natural line width to the distance $\De \om_j =
\om_{j+1} - \om_j \simeq 1/2\,\mu T$  between the lines is very
small numerically:
\beq\label{rat}
\frac{\Ga_j}{\De \om_j}\, \simeq
\frac{\mu^2}{16\pi^3}\,(j+1/2)(\sqrt{j(j+1)}\,\exp\left(-\mu\sqrt{j(j+1)}\right)\;{\lsim}\;
10^{-3}.
\eeq
Thus, the radiation spectrum of an isolated black hole is really
discrete.

At last, the total radiation intensity of a black hole is
\beq\label{itot}
I = \sum_j I_j = 0.150 A \,T^4.
\eeq
The numerical coefficient in this expression is close to that in
the total intensity of the common thermal radiation, i.e. to the
Stefan-Boltzmann constant which is $\pi^2/60 =0.164$. The point is
that the Rayleigh-Jeans contribution to the total intensity, which
is completely absent in the present spectrum, would be small
anyway.

Formulae (\ref{ija}) and (\ref{itot}) describe the thermal
radiation not only of bosons, photons and gravitons. They describe
as well the thermal radiation of fermions, massless neutrinos.
However, in the last case a proper account for the number of the
polarization states is necessary: for a two-component Dirac
neutrino the numerical factors in formulae (\ref{ija}) and
(\ref{itot}) will be two times smaller.

In fact, it was argued long ago~\cite{bem} that the discrete
thermal radiation spectrum of a black hole, with the equidistant
quantization rule for the horizon area, should fit the Wien
profile.

On the other hand, our conclusion of the discrete radiation
spectrum of a black hole in LQG differs drastically from that
of~\cite{bcr} according to which the black hole spectrum in LQG is
dense.

As to the nonthermal radiation of extremal black holes, described
by the terms with $\De J$ and $\De q$ in formulae (\ref{om}) and
(\ref{omc}), these effects are due to tunneling (see relatively
recent discussion of the subject, as well as detailed list of
relevant references, in~\cite{khr,kk1}). The loss of a charge by a
charged black hole is caused in fact by the Coulomb repulsion
between the black hole and emitted particles with the same sign of
charge. For a rotating black hole the reason is the interaction of
angular momenta: particles (massless mainly), whose total angular
momentum is parallel to that of a black hole, are repelled from
it.

\section*{Acknowledgements}
I appreciate numerous useful discussions with O.P. Sushkov. I am
grateful also to J. Bekenstein for the correspondence; in
particular, he has attracted my attention to the limit (\ref{da}).

\noindent An essential part of this work was done during my visit
to School of Physics, University of New South Wales, Sydney; I
wish to thank UNSW for the kind hospitality. The investigation was
supported in part by the Russian Foundation for Basic Research
through Grant No. 03-02-17612.

\end{document}